\documentclass[letterpaper,10pt]{IEEEtran}
\usepackage{amsmath,amsfonts}
\usepackage[colorinlistoftodos,prependcaption,textsize=footnotesize]{todonotes}
\usepackage{soul}
\newcommand{\revision}[1]{{\color{black}#1}}
\usepackage[switch]{lineno}
\usepackage{refcount}
\newcommand\scalemath[2]{\scalebox{#1}{\mbox{\ensuremath{\displaystyle #2}}}}
\usepackage{algorithm}
\usepackage{algpseudocode}
\usepackage{amsthm}
\newtheorem{definition}{Definition}
\usepackage{hyperref}
\usepackage[caption=false,font=normalsize,labelfont=sf,textfont=sf]{subfig}
\usepackage{graphicx}
\usepackage{float}
\usepackage{array}
\usepackage{todonotes}
\usepackage{placeins}
\usepackage{textcomp}
\usepackage{verbatim}
\usepackage{url}
\usepackage{cite}
\usepackage{stfloats}
\usepackage{siunitx}
\usepackage{xr}

\begin{document}

\title{Homotopy-Guided Potential Games for Congestion-Aware Navigation\thanks{Accepted at IEEE Robotics and Automation Letters (RA-L). DOI: \href{https://doi.org/10.1109/LRA.2026.3675907}{10.1109/LRA.2026.3675907}}}

\author{
  Mohammed I. I. S. Imran$^{1}$,
  Lasse Peters$^{\dagger,2}$,
  Michael Khayyat$^{\dagger,1}$,
  Stefano Arrigoni$^{1}$,\\
  Francesco Braghin$^{1}$,
  Laura Ferranti$^{2}$

\thanks{$^{\dagger}$These authors contributed equally as second authors.}
\thanks{$^{1}$M. I. I. Sathyamangalam Imran (corresponding author), M. Khayyat, S. Arrigoni, and F. Braghin are with the Department of Mechanical Engineering (DMEC), Politecnico di Milano, 20156 Milan, Italy.
\tt{\small
\url{mohammedirshadh.sathyamangalam@polimi.it};
\url{michael.khayyat@polimi.it};
\url{stefano.arrigoni@polimi.it};
\url{francesco.braghin@polimi.it}
}}
\thanks{$^{2}$L. Peters and L. Ferranti are with the Cognitive Robotics (CoR) Department, Delft University of Technology, 2628 CD Delft, The Netherlands.
{\tt\small\url{l.peters@tudelft.nl}; \url{l.ferranti@tudelft.nl}}}
\thanks{L. Ferranti is supported by the Office of Naval Research Global under Grant N62909-25-1-2027 (Project SECURE).}
}

\maketitle

\begin{abstract}
We address the multi-agent motion planning problem where interactions, collisions, and congestion co-exist. Conventional game-theoretic planners capture interactions among agents but often converge to conservative, congested equilibria. Homotopy planners, on the other hand, can explore topologically distinct paths, but lack mechanisms to account for the interdependence of agents' future actions. We propose a unified framework that leverages homotopy classes as structured strategy sets within a receding-horizon \revision{setup.} At each planning stage, a deterministic homotopy planner generates topologically distinct paths for each agent, conditioned on the joint configuration. To avoid intractable growth of candidate paths, we propose a simple heuristic filtering step that selects a top-$K$ subset of the most suitable congestion-free joint strategies to ensure computational tractability. These serve as initializations for a potential game that enforces homotopy-consistent constraints and yields a generalized open-loop Nash equilibrium (OLNE), with penalties discouraging abrupt strategy shifts in a receding-horizon setting. Simulations with three agents demonstrate improved efficiency (faster completion) and enhanced safety (greater inter-agent clearance, leading to reduced congestion) compared to a local baseline \revision{and NH-ORCA} that do not reason about homotopies. Hardware trials with two robots and one human demonstrate robustness to irrational behaviors, where our method adapts by switching to alternative feasible equilibria while the baseline game fails.
\end{abstract}


\section{Introduction} 
\label{sec:intro}
\begin{figure}[t]
\centering
\includegraphics[width=1\linewidth]{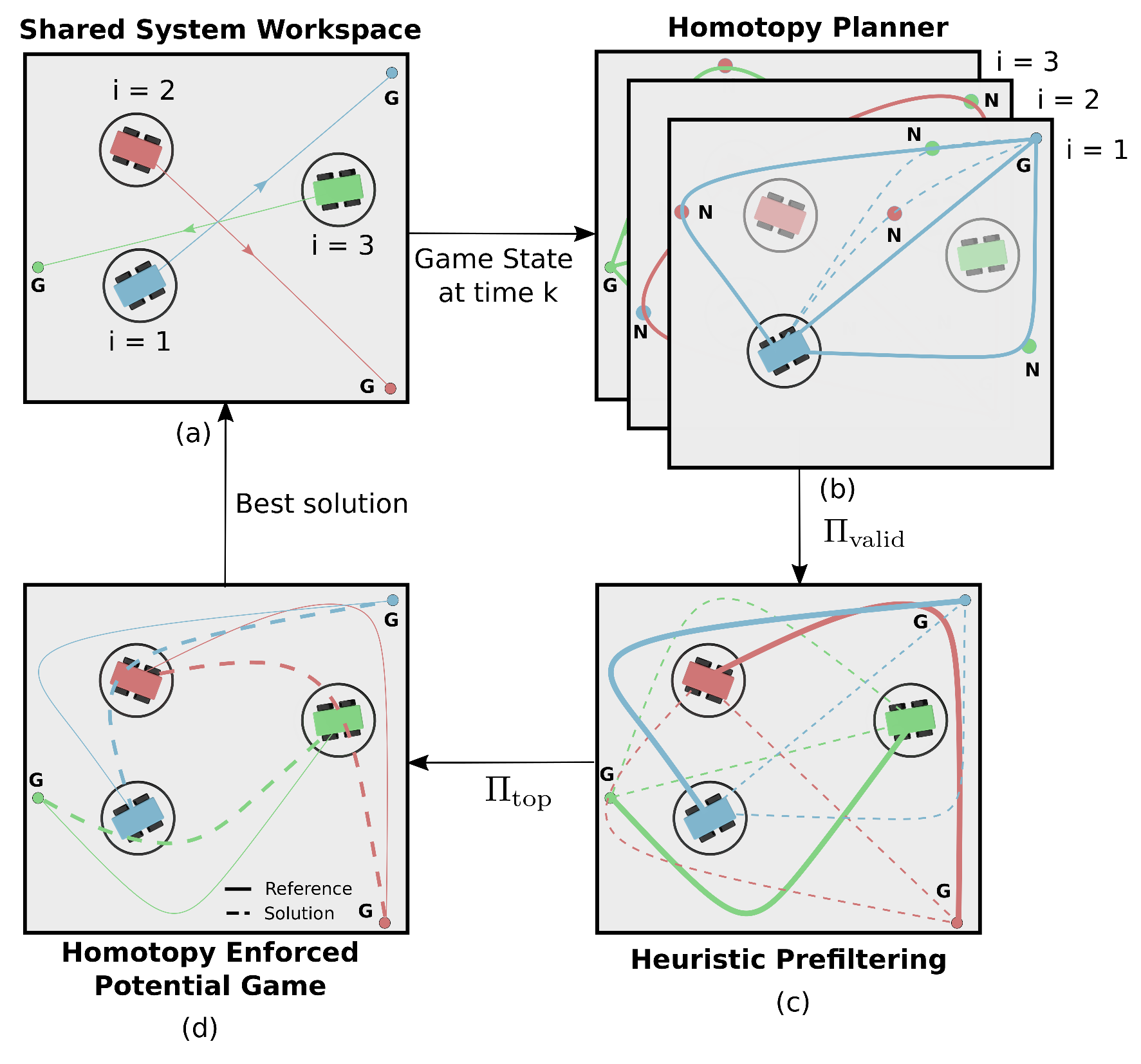}

\caption{Overview of the proposed 3-agent game framework. (a) Starting from a joint initial configuration with agent goals (G), (b) a homotopy planner generates topologically distinct paths using deterministic node sampling. Retained paths (solid) form valid joint homotopy sets $\Pi_{\text{valid}}$, while others (dotted) are filtered out. (c) A heuristic prefilter selects the top-$K$ congestion-free combinations $\Pi_{\text{top}}$, one of which is shown (solid line). (d) These serve as initializations for a homotopy-enforcing potential game, solved in receding-horizon fashion with regularization for consistent strategy selection.}
\label{fig_illustrator}
\end{figure}

% revision part is repsonsible for red.
Multi-Agent motion planning in dynamic environments poses significant challenges, as agents must navigate safely while anticipating and responding to others. Several paradigms address this complexity. Reactive methods, such as ORCA (Optimal Reciprocal Collision Avoidance) \revision{\cite{AlonsoMora2010OptimalRC}}, enable fast collision avoidance by assuming constant velocities for nearby agents, but often cause myopic and oscillatory behavior. Optimization-based approaches, such as distributed model predictive controller (MPC) \cite{Firoozi2024DistributedCoordination}, offer systematic constraint handling but rely on cooperation, limiting performance in adversarial and non-cooperative settings. Learning-based methods capture complex behaviors but depend on large datasets and offer limited safety guarantees \cite{b3}. 
%\linelabel{land}\revision{
%Game-theoretic approaches effectively model interactions but, for tractability, rely on local optimization that cannot overcome high-cost barriers between multiple Nash Equilibria (NEs) 
Game-theoretic approaches effectively model interactions but, for tractability, seek local equilibrium, which is limited not only by high-cost barriers but also by the local geometry of constraints between multiple Nash Equilibria (NEs) \cite{Peters2020InferenceBasedSA}, leading to congestion or overly cautious behavior. Homotopy-based planners generate topologically diverse paths, but do not reason about the interdependence of actions. To address these limitations, we propose a unified framework combining homotopy planning with potential games, enabling agents to plan trajectories that are both efficient and responsive to others in complex environments. An overview is shown in Fig.~\ref{fig_illustrator}.
\subsection{Related Work} 
\label{sec:relatedwork}
\IEEEpubidadjcol

\subsubsection*{Game-Theoretic Models}
Non-cooperative game theory provides a principled framework for modeling strategic interactions\cite{b18}. It has proven effective in adversarial settings such as competitive robotics\cite{b20}, as well as in non-adversarial coordination tasks including autonomous driving applications\cite{b19}.
Solution concepts for solving games include Stackelberg strategies \cite{simaan1973stackelberg} for asymmetric leader-follower roles, while %Nash Equilibrium
\revision{ NE} remains the standard for symmetric interactions \cite{leyton2022essentials}. The Generalized Nash Equilibrium Problem (GNEP) extends NE to include shared and individual constraints \cite{facchinei2010generalized}, and a key subclass is the Generalized Potential Game (GPG), governed by a global potential function \cite{monderer1996potential}. \\ \noindent Due to the often non-convex nature of these problems, computing global NE solutions is intractable \cite{conitzer2008complexity}. In practice, local approximations are used, with open-loop NE (OLNE) strategies preferred over feedback-based methods for their lower computational cost\cite{li2023costinference}. However, local NE solutions may produce congestion or overly cautious behavior, a limitation partially addressed by factorizing the game to decompose the interactions across spatio-temporal dimensions~\cite{9981416}.
\subsubsection*{Homotopy-based Planners}

Topological motion planning, on the other hand, offers robustness in environments with complex obstacle topologies from a single-agent perspective~\cite{b9}. 
Rather than optimizing solely for geometric efficiency, these methods exploit the configuration space’s topology by categorizing trajectories into \textit{homotopy classes}: equivalence classes of obstacle-avoiding, continuously deformable paths \cite{b10}. A key distinction exists between \textit{fixed-end homotopy}, which preserves both start and goal positions, and \textit{free-end homotopy}, which allows flexibility within terminal regions \cite{b12}. Compact representations such as \textit{$H$-signatures}~\cite{b13} and \textit{winding numbers} \cite{b14} are widely used to encode and distinguish homotopy classes. H-signatures record signed intersections with reference rays from obstacle features, while winding numbers measure net encirclement around obstacles. These descriptors support exploration of topologically distinct paths and promote diversity in planning. Topological methods have proven effective in socially compliant navigation \cite{b15}, deadlock avoidance \cite{b16}, and robust autonomous vehicle planning~\cite{b17}. However, interactions among multiple decision-making agents are not explicitly captured in homotopy-based planning, limiting its ability to address multi-agent interactions.\\
Despite their complementary strengths, topological planning and game-theoretic modeling are rarely integrated. 
%\linelabel{bin}\revision{
A notable exception is \cite{Khayyat2024TacticalGD}, which treats homotopy as a binary, pairwise decision variable defining the passing order at intersections of predefined agent paths, assuming perfect path following. Our work considers all agents jointly, combining homotopy planning with game-theoretic reasoning to capture path diversity and strategic interactions.
%} \linelabel{bin_end}
\begin{figure}[ht]
\centering
\includegraphics[trim={0 2.8cm 0 2cm},clip,  width=0.45\linewidth]{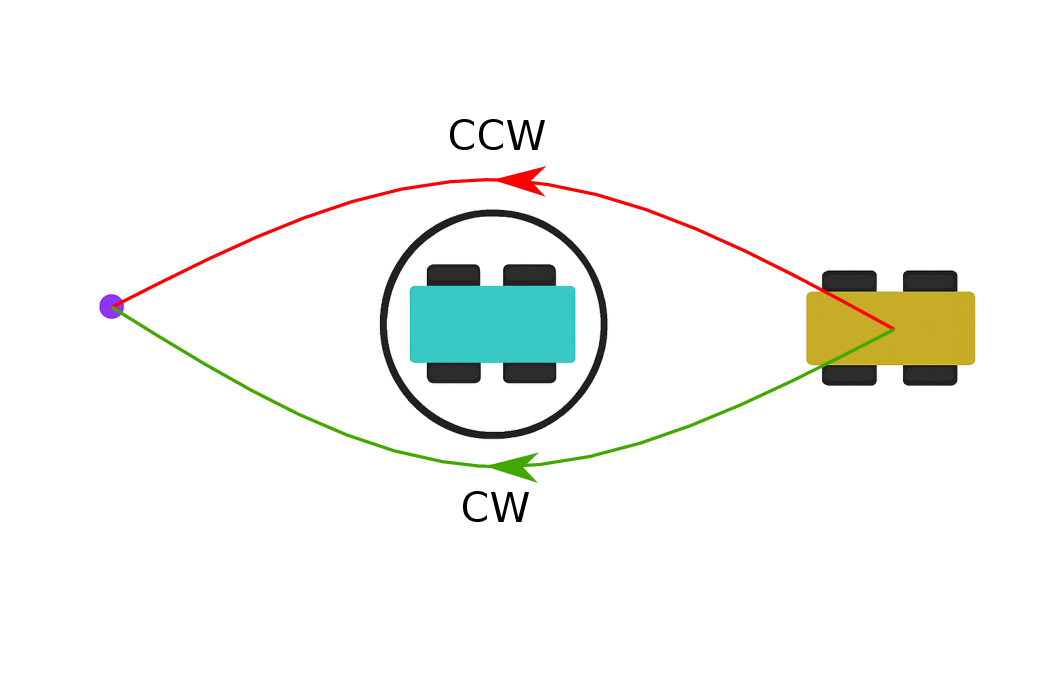}
\caption{An example of fixed-end homotopes that could be generated for the yellow agent around the blue agent to reach a designated goal (Purple).}
\label{fig_ex_homo}
\end{figure}
\subsection{Contribution}
\label{sec:contri}
This letter makes three main contributions toward congestion-aware multi-agent navigation:
\begin{enumerate}
    \item \textbf{Homotopy-guided game formulation:} We introduce a non-cooperative potential game where agents select strategies from distinct homotopy classes, enabling both strategic coordination and topological path diversity.
    \item \textbf{Heuristic Prefiltering:} We design a lightweight prefiltering step that discards unsafe or congested joint homotopy combinations, ensuring tractable computation in multi-agent settings.
    \item \textbf{Regularized receding-horizon execution:} We develop a homotopy-consistent regularization that ensures consistent homotopy strategies across horizons, avoiding abrupt strategy shifts.
    %that stabilizes strategy evolution across horizons, by avoiding abrupt strategy shifts.
\end{enumerate}
\revision{We validate the framework in simulation and hardware experiments. In simulation, our approach demonstrates improved efficiency (faster task completion) and safety (increased average inter-agent distance) compared to baseline game and NH-ORCA that do not leverage homotopies. In hardware, it shows robustness to an irrational agent, compared to a baseline potential game that does not reason about homotopy.}

\section{Preliminaries} 
\label{sec:prelims}

\subsection{Homotopy Planning}
\label{sec:homotopy}
Consider a system of \( N \) agents operating in a shared workspace \( \mathcal{W} \subset \mathbb{R}^2 \). For each agent \( i \in \mathcal{A} := \{1, \dots, N\} \), the remaining agents \( \mathcal{A} \setminus \{i\} \) are modeled as static circular obstacles
\footnote{We assume static obstacles during homotopy planning, but we relax this assumption during joint strategy synthesis, where homotopic path referencing and joint-homotopy-enforcing constraints adds dynamic obstacle behavior.}. Let \( \mathcal{O}_i = \{ j \mid j \in \mathcal{A} \setminus \{i\} \} \) denote the set of obstacles for agent \( i \), with cardinality \( |\mathcal{O}_i| = N-1 \). Each obstacle \( j \in \mathcal{O}_i \) induces a closed disk region \( R_j \subset \mathcal{W} \), centered at the current position of agent \( j \), and the total occupied space is defined as \( \mathcal{R}_i := \bigcup_{j \in \mathcal{O}_i} R_j \).
A continuous path \( \hat{\tau}_i: [0,1] \rightarrow \mathcal{W} \setminus \mathcal{R}_i \) is said to belong to a homotopy class if it can be continuously deformed into another path with the same endpoints without intersecting any obstacle~\cite{b13}. Fig.~\ref{fig_ex_homo} provides an example.
\begin{definition}[$H$-signature]
\label{definition_hsign}
Given a continuous path \( \hat{\tau}_i \) for agent \( i \), the \textit{$H$-signature}, \( \mathcal{H}(\hat{\tau}_i) \), is a vector-valued topological descriptor defined as
\begin{equation}
\label{eq:homotopy_signature}
\mathcal{H}(\hat{\tau}_i) = \left[ \mathcal{H}_1(\hat{\tau}_i), \dots, \mathcal{H}_{N-1}(\hat{\tau}_i) \right] \in \{\text{CCW}, \text{CW}, \text{S}\}^{N-1},
\end{equation}
where each \( \mathcal{H}_j(\hat{\tau}_i) \) denotes the winding direction of the path around obstacle \( O_j \):  
\textnormal{CCW} for counterclockwise, \textnormal{CW} for clockwise, and \textnormal{S} for straight if the obstacle is not encircled.
\end{definition}
\noindent For implementation purposes, a discrete path $\tilde{\tau}_i$ is obtained by discretizing $\hat{\tau}_i$ along arc-length using a reference velocity and a discretization time-step. Although the homotopy class is defined for continuous paths, this discretization preserves the topological properties of the path. The winding behavior around obstacles remains unchanged, ensuring \( \mathcal{H}(\tilde{\tau}_i) = \mathcal{H}(\hat{\tau}_i) \).

\subsection{Game Theory}
\label{sec:gametheory}
To capture strategic interactions among agents traversing a shared workspace, we model motion planning as a finite-horizon dynamic noncooperative game over \(T\) steps. Let the set of rational agents be \(\mathcal{A}=\{1,\dots,N\}\). Each agent \(i\in\mathcal{A}\) chooses a control sequence \(u^i=\{u_k^i\}_{k=0}^{T-1}\) that results in a state trajectory \(x^i=\{x_k^i\}_{k=0}^{T}\) through the discrete-time equations governing the dynamics:
\begin{equation}
\label{eq:dynamics}
    x_{k+1}^i = f^i(x_k^i, u_k^i), \qquad k=0,\dots,T-1.
\end{equation}
We collect agent \(i\)'s decision variables as \(z^i := (x^i, u^i)\), and adopt the shorthand \(\square^{-i}\) to denote the collection of the quantity \(\square\) for all players except \(i\), i.e., those in \(\mathcal{A}\setminus\{i\}\). Since the agents traverse a common workspace, then the feasible decisions of agent \(i\) depend on those of other agents via
\begin{equation}
    z^i \in \mathcal{Z}^i(z^{-i}),
\end{equation}
with
\begin{equation}
\begin{split}
    \mathcal{Z}^i(z^{-i}) = \{\, z^i=(x^i, u^i)\ \big|\ & g^i_{\mathrm{i}}(z^i, z^{-i}) \le 0,\\
    & g^i_{\mathrm{e}}(z^i, z^{-i}) = 0 \,\},
\end{split}
\end{equation}
where \(g^i_{\mathrm{i}}\) and \(g^i_{\mathrm{e}}\) collect inequality constraints (e.g., collision-avoidance, state bounds, and actuator limits), and equality constraints (e.g., dynamics \eqref{eq:dynamics}, initial conditions), respectively.\\
Player $i$ seeks $z^i \in \mathcal{Z}^i(z^{-i})$ that minimizes the cost $J^i(z^i, z^{-i})$; Agent $i$'s trajectory-optimization problem is
\begin{equation}
\label{eq:gnep}
P^i \left\{
\begin{split}
&S(z^{-i}):=\arg\min_{z^i}\ \ J^i(z^i, z^{-i})\\[0.5ex]
&\qquad\qquad\qquad\text{s.t.}\ \ z^i \in \mathcal{Z}^i(z^{-i})
\end{split}
\right.
\end{equation}
Then, the game can be defined as follows:
\begin{definition}[\textit{Game}]
A dynamic game is the tuple \(\mathcal{G} = \langle \mathcal{A}, \mathcal{Z}, \{J^i\}_{i\in\mathcal{A}}\rangle\), where the joint feasible set is
\begin{equation*}
    \mathcal{Z} = \big\{\, z=(z^1,\dots,z^N)\ \big|\ z^i \in \mathcal{Z}^i(z^{-i})\ \ \forall i\in\mathcal{A} \,\big\}.
\end{equation*}
\end{definition}

Let \(S^i(z^{-i})\) denote the solution set of \(P^i\) given \(z^{-i}\). We can now state the following definition:

\begin{definition}[\textit{Generalized Nash Equilibrium Problem (GNEP)}]
    Find \(\hat z\) such that \(\hat z^i \in S^i(\hat z^{-i})\) for all \(i \in \mathcal{A}\).
\end{definition}

Such a strategy profile, $\hat{z}$, is called a generalized NE, and it represents a joint strategy set where no agent can decrease its own cost by unilaterally perturbing its strategy in a locally feasible direction.
Computing an NE of \eqref{eq:gnep} is generally challenging because players' decisions are coupled through both costs and constraints. We exploit additional problem-specific structure in \(J^i\). A useful case is that of \textit{generalized potential games} in which unilateral cost differences coincide with differences of a single scalar potential; in such games, NE search reduces to optimizing that potential function~\cite{sagratella2017algorithms, monderer1996potential}. \cite{monderer1996potential} shows that these minimizers can be viewed as a refinement of the NE set~\cite{carbonell2014refinements}, and are favorable from a social, or collective, perspective.

\begin{definition}[\textit{Generalized Potential Game}]
\label{def:wpg}
A GNEP is a generalized potential game if there exists a function \(F:\mathcal{Z}\to\mathbb{R}\) such that, for all \(i\in\mathcal{A}\),
\begin{equation*}
\begin{split}
    J^i(z^{i}, z^{-i}) - J^i(\hat z^{i}, z^{-i})
    = F(z^{i}, z^{-i}) - F(\hat z^{i}, z^{-i}),\\
    \forall\, z^{i}, \hat z^{i} \in \mathcal{Z}^i(z^{-i}),\ \ z^{-i}\in \mathcal{Z}^{-i}.
\end{split}
\end{equation*}
In this case, \(\mathcal{G} = \langle \mathcal{A}, \mathcal{Z}, \{J^i\}_{i\in\mathcal{A}} \rangle\) and
\(\mathcal{G}^\dag = \langle \mathcal{A}, \mathcal{Z}, \{F\}_{i\in\mathcal{A}} \rangle\) share the same set of Nash equilibria.
\end{definition}
\begin{algorithm}[t]
\caption{Homotopy Planner}
\label{homo_planner_algo}
\begin{algorithmic}[1]
\footnotesize
\State \textbf{Input:} Agent set $\mathcal{A}$ with positions $\{p_i\}_{i \in \mathcal{A}}$, initial and goal positions $x_0^i$, $x_{\text{goal}}^i$, sampling distance $\delta$, agent radii \(r\), B-spline parameters $(S_{sp}, D_{sp})$, reference velocity $v_{\text{ref}}^i$, timestep $\delta t$
\State \textbf{Output:} Set of valid, joint topologically distinct paths $\Pi_{\text{valid}}$
\ForAll{$i \in \mathcal{A}$}
    \State Initialize graph nodes $V^i \gets \{x_0^i\}$ and edges $E^i \gets \emptyset$
 %   \ForAll{$j \in \mathcal{A} \setminus \{i\}$}
        \State Sample candidate nodes $Q_{ij,L}$,$Q_{ij,R}$ around $j \in \mathcal{A} \setminus \{i\}$ at $\delta$ along perpendicular directions $\mathbf{d}_{ij}^{\perp,L/R}$ to the unit vector $d_{ij}$ from \(p_i\) to \(p_j\)
        %\State Compute unit vector $\mathbf{d}_{ij}$ from $i$ to $j$
        %\State Compute perpendiculars $\mathbf{d}_{ij}^{\perp,L}$, $\mathbf{d}_{ij}^{\perp,R}$ from $\mathbf{d}_{ij}$ at $p_j$
        %\State Sample candidate points $Q_{j,L}, Q_{j,R}$ along perpendiculars at $\delta$
        \State For each $Q \in \{Q_{ij,L}, Q_{ij,R}\}$: if conflicts with other agents $k \in \mathcal{A} \setminus \{i,j\}$, use midpoint fallback $Q_{ij,\text{alt},L/R}$ if valid, else mark $j,k$ as composite obstacle; otherwise, add $Q$ to $V^i$
 %   \EndFor
    \State Connect nodes $V^i$ with edges $E^i$ that avoid agent disks of radius \(r\)
    \State Construct graph $\mathcal{G}^i = (V^i, E^i)$
    \State Enumerate candidate paths $\hat{\Pi}^i$ from $x_0^i$ to $x_{\text{goal}}^i$
    \State Group $\hat{\Pi}^i$ by H-signature: 
$\Gamma^i[\mathcal{H}] = \{\hat{\tau}^i \in \hat{\Pi}^i \mid \mathcal{H}(\hat{\tau}^i) = \mathcal{H}\}$

    \ForAll{$\mathcal{H} \in \Gamma^i$}
        \State Select shortest path: $\hat{\tau}_{\min}^i(\mathcal{H}) \gets \arg\min_{\hat{\tau}^i \in \Gamma^i[\mathcal{H}]} \text{Length}(\hat{\tau}^i)$
        %\State If $\hat{\tau}_{\min}^i(\mathcal{H})$ is valid, smooth path: $\tilde{\tau}^i \gets \text{BSpline}(\hat{\tau}_{\min}^i(\mathcal{H}); S_{sp}, D_{sp})$
        \State Smooth path $\hat{\tau}_{s}^i(u) \gets \text{BSpline}(\hat{\tau}_{\min}^i(\mathcal{H}); S_{sp}, D_{sp})$
        %\State Smooth path $\hat{\tau}_{s}^i(u) \gets \text{BSpline}(\hat{\tau}_{\min}^i(\mathcal{H}); S_{sp}, D_{sp})$
        \State Discretize path: $\tilde{\tau}^i \gets \text{Discretize}(\hat{\tau}_{s}^i, s_{\text{ref}}^i), \quad s_{\text{ref}}^i = v_{\text{ref}}^i \cdot \delta t$
%\State Discretize path $\tilde{\tau}^i$ along arc length $s_{\text{ref}}^i = v_{\text{ref}}^i \cdot \delta t$
        \State Add $\tilde{\tau}^i$ to $\Pi_{\text{valid}}^i$
        %\If{$\hat{\tau}_{\min}^i(\mathcal{H})$ valid}
        %    \State Smooth path: $\hat{\tau}_{s}^i(u) \gets \text{BSpline}(\hat{\tau}_{\min}^i(\mathcal{H}); S_{sp}, D_{sp})$
        %    \State Discretize using arc length $s_{\text{ref}}^i = v_{\text{ref}}^i \cdot \delta t$: $\tilde{\tau}^i \gets \hat{\tau}_{s}^i(u)$
         %   \State Add discretized path $\tilde{\tau}^i$ to $\Pi_{\text{valid}}^i$  
        %\EndIf
    \EndFor
\EndFor
\State \Return{$\Pi_{\text{valid}} = \{ (\tilde{\tau}^1, \ldots, \tilde{\tau}^{N}) \mid \tilde{\tau}^i \in \Pi_{\text{valid}}^i \}$}
\end{algorithmic}
\end{algorithm}
\section{Interactive Homotopy-Aware Planning}
\label{sec:methodology}
This section presents our main contribution: a homotopy-aware game-theoretic motion planner.
Our approach consists of three main stages, as illustrated in Fig.~\ref{fig_illustrator}, each of which we detail below: (\romannumeral1) generating a set of high-level candidate plans via homotopy planning (Fig.~\ref{fig_illustrator}a, Section~\ref{sec:homotopyplanner}), (\romannumeral2) reducing the number of candidate plans to a small number of promising options (Fig.~\ref{fig_illustrator}b, Section~\ref{sec:scalarranking}), and (\romannumeral3) refining the most promising high-level strategy into interaction-aware low-level strategies via homotopy-informed game-theoretic planning (Fig.~\ref{fig_illustrator}c, Section~\ref{sec:game-refinement}).

\subsection{Homotopy Planner}
\label{sec:homotopyplanner}
\begin{figure*}
    \centering
    \begin{minipage}{1\textwidth}
        \centering
        \includegraphics[width=1\linewidth]{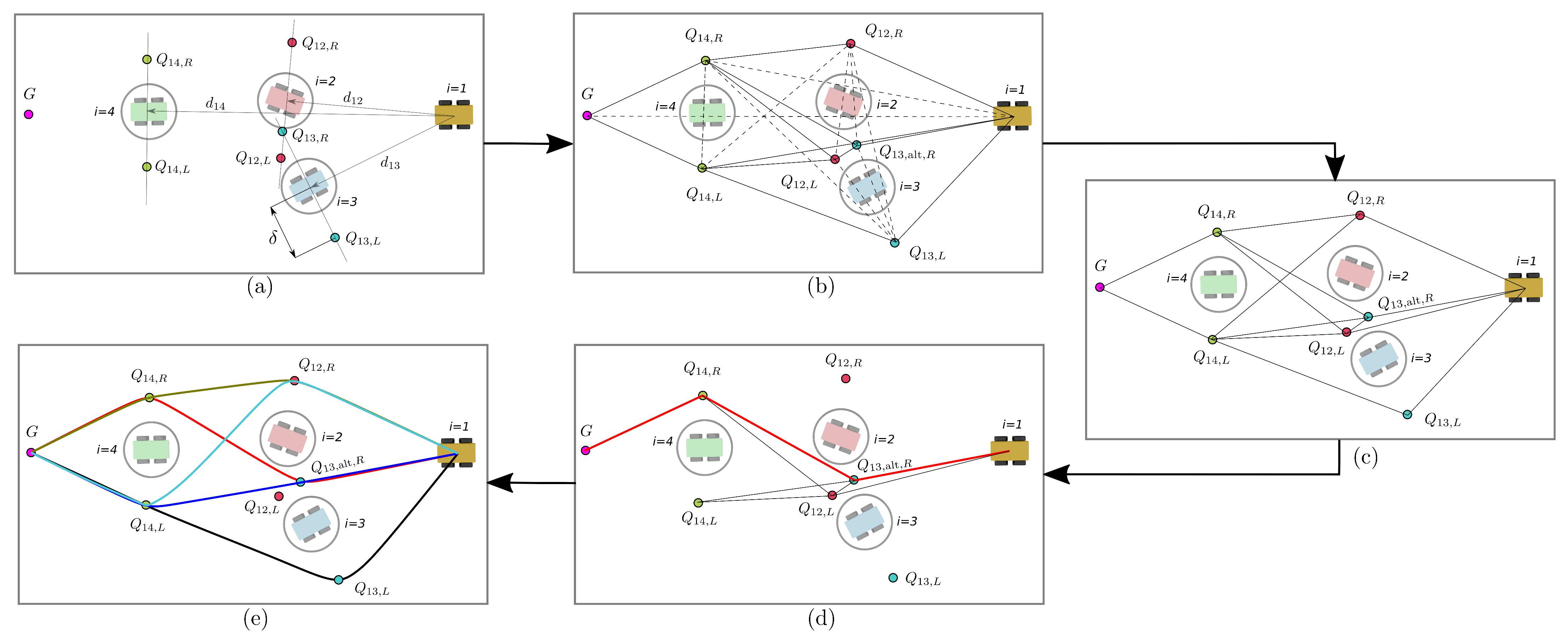}
        \caption{\revision{Schematic illustration of graph construction for the yellow agent ($i=1$) towards its Goal \(G\) by the homotopy planner with respect to other agents \(i={2,3,4}\) represented by red, blue and green respectively: (a) node placement, (b) edge validation with node adjustment ($Q_{13,\text{alt},R}$), (c) graph construction, (d) pruning redundant paths within same homotopy class, (e) final set of topologically distinct homotopy paths.}}
        \label{fig:homotopy_full_explain_4}
    \end{minipage}
\end{figure*}

Algorithm~\ref{homo_planner_algo} summarizes our homotopy planner, \revision{illustrated schematically in Figure~\ref{fig:homotopy_full_explain_4} for a single agent among others}. The planner is integrated with a potential game framework to address limitations of multiple local \revision{NEs} in GPGs, where certain solutions may lead to undesirable outcomes such as congestion or overly cautious behavior. \revision{
%\linelabel{sec_1_no}
To overcome these challenges, our method leverages homotopy reasoning in a game-theoretic setting to divide the solution space into distinct topological classes. This enables a more comprehensive search for efficient local NE.\\
%\linelabel{sec_1_no_end}
}\noindent Like many global path planning approaches, our method ultimately performs graph search. The key distinction lies in graph construction: we use $H$-signatures (Definition~\ref{definition_hsign}) to deterministically place nodes, structuring the combinatorial search over homotopy classes. This design allows explicit enumeration of all homotopy classes, ensuring topological completeness. Although the complexity remains combinatorial, focusing on representative classes reduces redundant exploration and significantly improves planning efficiency.\\Despite these advantages, scalability remains a challenge. The graph size grows with the number of agents, potentially leading to a combinatorial explosion. As a result, we focus on games with a moderate number of agents (typically 3--5). Strategies for improving scalability in larger multi-agent systems can be found in~\cite{10341677}.
\\\noindent\textbf{Planner Workflow.} 
For each agent \( i \), the node set is initialized with its start position (Step~4). Candidate nodes are sampled by applying perpendicular offsets from the line connecting the agent’s start position to each other agent \revision{(Step~5) and Figure~\ref{fig:homotopy_full_explain_4}(a)}. Conflicts are resolved via midpoint insertion or by marking composite obstacles \revision{(Step~6) and Figure~\ref{fig:homotopy_full_explain_4}(b)}. The resulting valid nodes and edges form a graph \( \mathcal{G}^i \) \revision{(Steps~7--8) and  Figure~\ref{fig:homotopy_full_explain_4}(c)}, from which continuous paths \( \hat{\tau}^i \in \hat{\Pi}^i \) to the goal are enumerated \revision{(Step~9)}. These are grouped by their \( H \)-signatures \revision{(Step~10)}, and the shortest valid representative per class is selected \revision{(Step~12) and Figure~\ref{fig:homotopy_full_explain_4}(d)}. Each selected path is then smoothed using a B-spline and discretized along arc length based on the agent’s reference velocity and timestep to yield \( \tilde{\tau}^i \) \revision{(Steps~13--15) and  Figure~\ref{fig:homotopy_full_explain_4}(e)}. The final joint homotopy set \( \Pi_{\text{valid}} \) is constructed as the Cartesian product across all agents (Step~18).
\\\noindent\textbf{Remark 1.}  Path planners such as Probabilistic Road Mapping (PRM)~\cite{prm_article} and Rapidly-exploring Random Trees (RRT)~\cite{Karaman2011} also perform graph search, but construct the graph via random sampling. This allows control over graph size through the number of samples, helping to avoid combinatorial explosion. However, sampling-based methods may require many samples to discover all homotopy classes. Our deterministic approach avoids this by explicitly structuring the graph around homotopy classes, which ensures topological completeness but scales with the number of agents~\cite{Janson2018}.

\subsection{Heuristic Prefiltering}
\label{sec:scalarranking}
\revision{
%\linelabel{sec_II_nov}\linelabel{homo_gr}
The number of multi-agent homotopy combinations produced by the homotopy planner grows exponentially with the number of agents \( N \), assuming each agent has \( M(N) \) valid homotopy paths. This yields \( O((M(N))^N) \) total joint combinations in the worst case. This rapid growth renders exhaustive evaluation of the full set \( \Pi_{\text{valid}} \) computationally intractable.
To address this, we score each combination based on a set of joint metrics such as collision risk, path length, and path smoothness, retaining only the top-$K$ combinations and thereby reducing the search space. %\linelabel{sec_II_nov_end}
}
Algorithm~\ref{alg:scalar_filtering_graph} outlines this prefiltering procedure.\\ 

\begin{figure}[ht]
\centering
\includegraphics[width=1\linewidth]{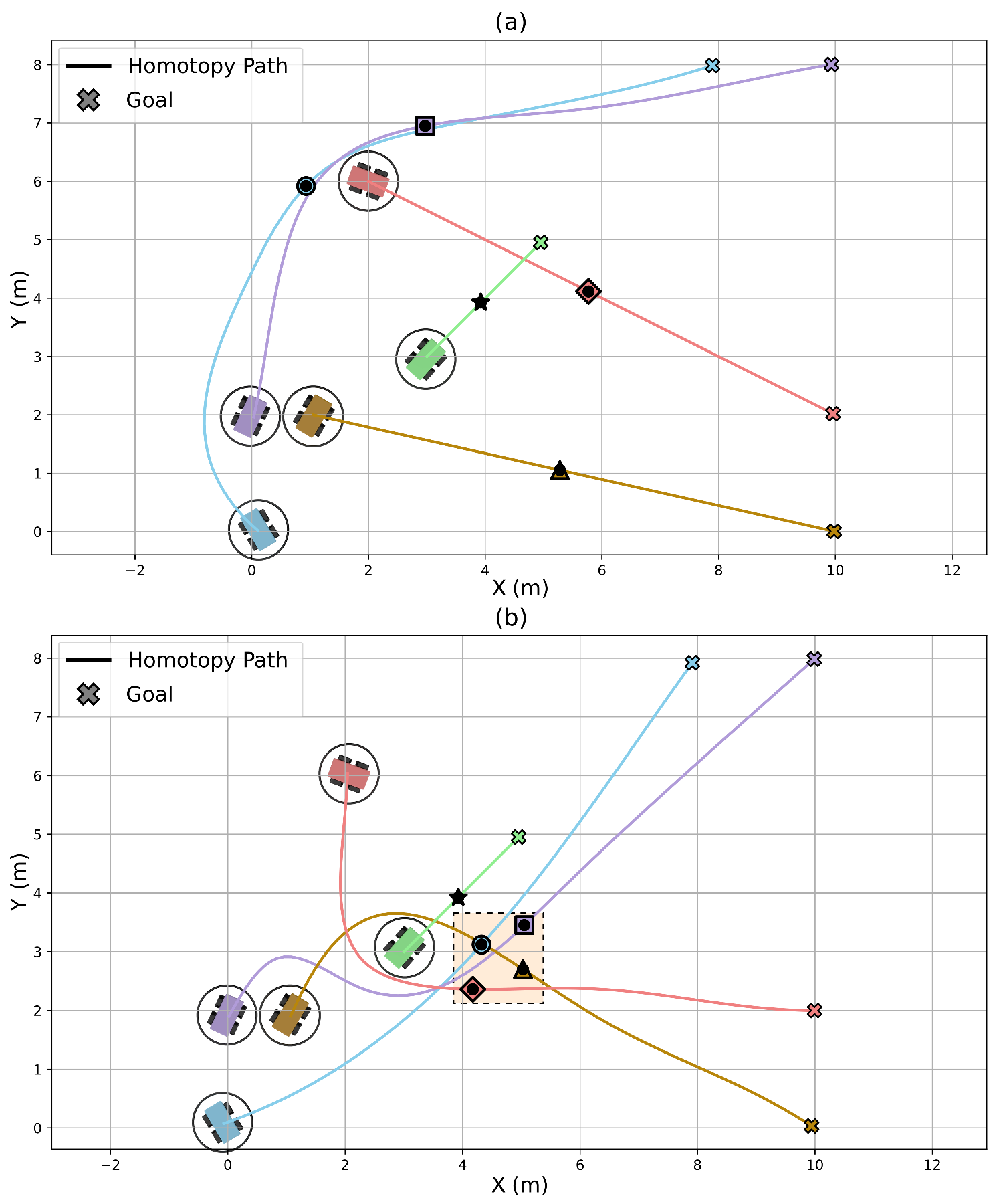}
\caption{Ranking comparison between two homotopy combinations, with agents assigned as follows: blue: Agent 1, violet: Agent 2, yellow: Agent 3, red: Agent 4, and green: Agent 5. Symbols along each path indicate predicted positions at \(s = 20\) steps%into the future
, shown for illustration. Joint paths from homotopy planner highlight congestion near predicted positions. (a) Best homotopy combination (lowest score): agents are well-separated. (b) Worst homotopy combination (highest score): bounding box highlights spatial clustering of Agents 1–4 at \(s = 20\), indicating high congestion.}
\label{fig_com}
\end{figure}

\noindent\textbf{Prefiltering Mechanism.} For each candidate joint homotopy $C \in \Pi_{\text{valid}}$, agent paths are re-sampled to compute length and smoothness penalties \revision{(Step~5--6)}. At each re-sample index, a proximity graph is built \revision{(Steps~7--8)} to penalize agents that are too close. The total score combines length, smoothness, and proximity terms \revision{(Step~9)}, and all combinations are ranked accordingly \revision{(Step~10)}. Finally, the top-$K$ lowest-scoring joint strategies $\Pi_{\text{top}}$ are retained \revision{(Steps~12--13)}. %for the subsequent potential game.
\\\noindent\textbf{Example.} Fig.~\ref{fig_com} illustrates a representative 5-agent scenario. The valid homotopy set comprises \( |\Pi_{\text{valid}}| = 270 \) multi-agent combinations, ranked using \( P = 40 \) \revision{
%\linelabel{p1}
(More points provide a finer evaluation but increase computational effort)
%\linelabel{p1_e}
}, \( \lambda_L = 1 \) \revision{
%\linelabel{p2}
(penalizes longer paths)
%\linelabel{p2_e}
}, \( \lambda_P = 2.0 \) \revision{
%\linelabel{p3}
(penalizes close agents)
%\linelabel{p3_e}
}, \( \lambda_S = 1 \) \revision{
%\linelabel{p4}
(penalizes curved trajectories)
%\linelabel{p4_e}
}, and a proximity threshold \( d_{\text{th}} = 1.5\,\text{m} \) \revision{
%\linelabel{p5}
(distance below which agents are considered too close)
%\linelabel{p5_e}
} in Algorithm~\ref{alg:scalar_filtering_graph}. For illustration, two combinations are compared at sampling index \( s = 20 \) \revision{
%\linelabel{p6_s}
(20 steps into the future)
%\linelabel{p6_se}
}. Although the ranking score accounts for all steps, index \( s \) is used solely for visualization. In the worst-ranked case, a bounding box of size \( d_{\text{th}} \) highlights clustering of Agents 1--4, indicating high congestion and a higher proximity penalty; the best-ranked combination maintains separation, yielding a lower score.
\begin{algorithm}[t]
\caption{Heuristic Prefiltering}
\label{alg:scalar_filtering_graph}
\begin{algorithmic}[1]
\footnotesize
\State \textbf{Input:} Multi-agent homotopy set $\Pi_{\text{valid}}$; resample size $P$; weights $\lambda_L, \lambda_P, \lambda_S$; distance threshold $d_\text{th}$; top-K $\text{top}_K$
\State \textbf{Output:} Best score $S^*$, top-K combinations $\Pi_{\text{top}}$
\State Initialize $S^* \gets \infty$, score list $\mathcal{S} \gets []$
\ForAll{\(\mathcal{C} = (\tilde{\tau}^1,\ldots,\tilde{\tau}^N) \in \Pi_{\text{valid}}\)}
    \State $\mathcal{R} \gets$ resample paths in $\mathcal{C}$ to $P$ points
    \State Update path length \(L\) and smoothness cost \(S_{\text{smooth}}\) from \(\mathcal{R}\)
    \State Construct proximity graph $G^t$ and connected components $C^t_k$ for $t = 1, \dots, P$ using $d_\text{th}$ and $\mathcal{R}$
    \State Compute proximity cost $S_{\text{prox}}$ (from $G^t$ with \(C^t_k\geq 2\))
    \State Compute total score $S \gets \lambda_L L + \lambda_P S_{\text{prox}} + \lambda_S S_{\text{smooth}}$
    \State Store $(\mathcal{C}, S)$ in $\mathcal{S}$; update $S^*$ if $S < S^*$
\EndFor
\State Sort \(\mathcal{S}\) by \(S\); select top-K combinations \(\Pi_{\text{top}}\)
\State \Return $S^*, \Pi_{\text{top}}$
\end{algorithmic}
\end{algorithm}
\subsection{Game-Theoretic Modelling}
\label{sec:game-refinement}

\begin{figure}
\centering
\includegraphics[width=1\linewidth]{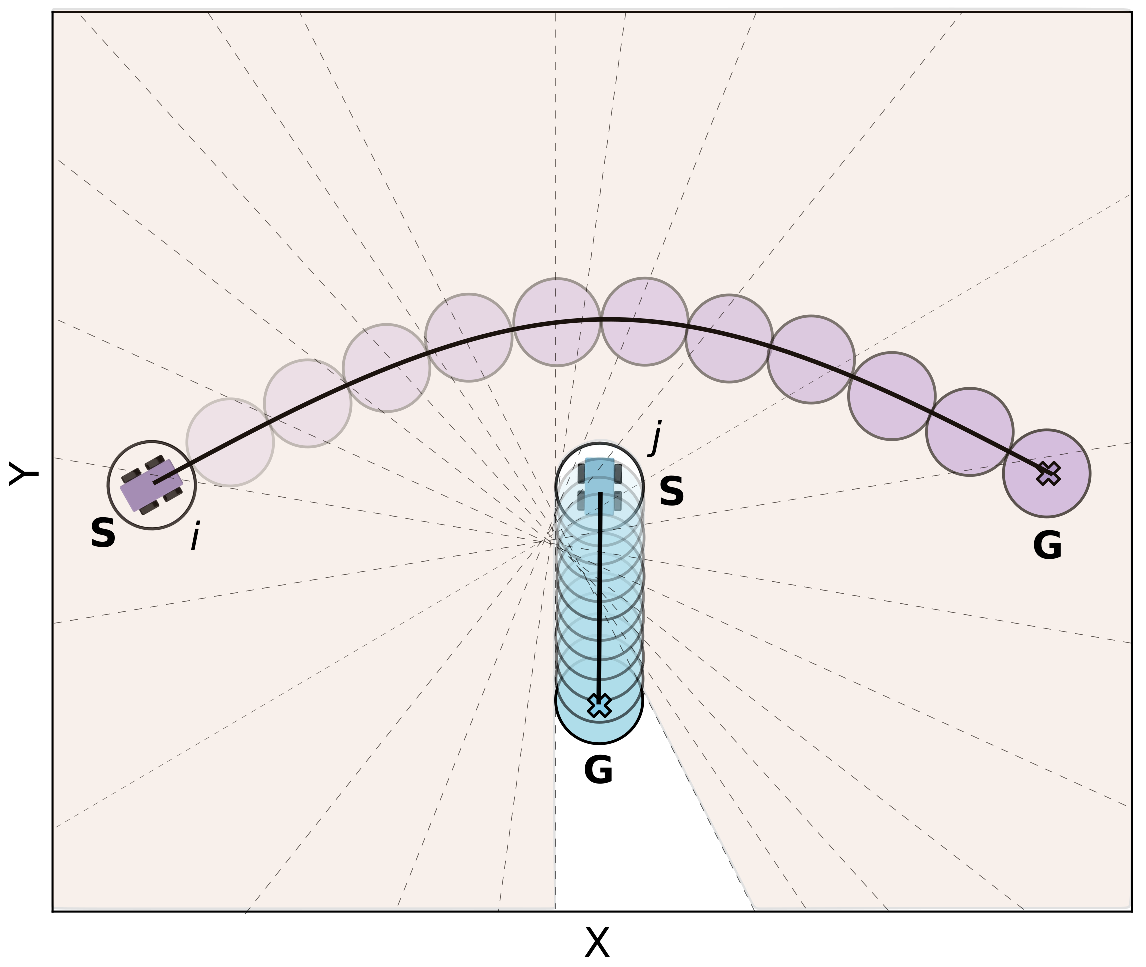}

\caption{Schematic illustration of the homotopy-enforcing constraint, a technique adapted from~\cite{oscar}, for Agent \(i\) (purple) influenced by Agent \(j\) (blue). ``S" and ``G" denote start and goal positions. Solid lines represent homotopy reference paths, while circles with increasing opacity indicate the expected agent positions along these paths, spaced by arc length according to their reference velocities. Dashed lines show separating hyperplanes defined by unit normal \(\hat{A}^{ij}_k\), pointing from \(i\) toward \(j\). The safety margin \(\beta_o r_o\), visualized as Agent \(j\)'s circle radius, defines the spatial influence region and ensures \(i\) stays on the correct side of each hyperplane, as required by the homotopy signature. Shaded regions show feasible space for \(i\) to follow its path while maintaining separation from \(j\).}
\label{fig_constraint}
\end{figure}

To exploit the filtered topological initializations $\boldsymbol{\Pi}_{\text{top}} = \{\tilde{\tau}_{\text{top}}^{1}, \dots, \tilde{\tau}_{\text{top}}^{K}\}$, where each $\tilde\tau_{\text{top}}^{m}$ with $m=1,\dots, K$ is a joint candidate path, we formulate an optimal control problem (OCP) whose solution corresponds to the $N$-player potential game solution over a $T$-step prediction horizon, as defined in Eq.~\eqref{potential_nom}. \revision{The potential function \(\Phi_\mathrm{h}\) aggregates the individual objectives 
\(J^i(u^i;x^i_\text{init},\tilde{\tau}^i_{\text{ref}})\) from Eq.~\eqref{individual_ocp_common_tau_1}, with \(\tilde{\tau}^i_{\text{ref}} \in \tilde\tau^m_\text{top}\). This aggregation aligns each agent's incentive, minimizing its deviations from its reference homotopy path \(\tilde{\tau}^i_{\text{ref}}\) and control effort, with the global objective such that any OLNE corresponds to a local minimum of \(\Phi_\mathrm{h}\)}. The OCP in \eqref{potential_nom} is solved in a receding-horizon (MPC) fashion in series for each \(\tilde{\tau}^m_\text{top} \in \Pi_\text{top} \):
\begin{equation}
\label{potential_nom}
\begin{aligned}
\noindent \min_{\mathbf{u}} & \overbrace{
\scalemath{0.85}{
\sum_{k=0}^{T-1} \sum_{i=1}^{N} \left( \| x^i_k - \tilde{\tau}^i_{k,\text{ref}} \|^2_{w^i_x} + \| u^i_k \|^2_{w^i_u} \right) + \sum_{i=1}^{N} \| x^i_T - \tilde{\tau}^i_{T,\text{ref}} \|^2_{w^i_T} + \lambda
}}^{\Phi_\mathrm{h}(\mathbf{u}; \mathbf{x}_\text{init}, \tilde{\tau}^m_{\text{top}}) :=} \\
\text{s.t.}\quad & 
\begin{array}{rcll}
\mathbf{x}_0 & = & \mathbf{x}_\text{init}, & \forall i \in \mathcal{A}, \\
h(\mathbf{x}_k, \mathbf{u}_k) & = & 0, & \forall k \in \{0,\dots, T{-}1\}, \\
g_\mathrm{h}(\mathbf{x}_k,  \tilde{\tau}^m_{\text{top}}) & \le & 0, & \forall k \in \{0,\dots, T{-}1\}, \tilde{\tau}^m_\text{top} \in \Pi_{\text{top}}, \\ 
\text{u}_\text{L} \leq \mathbf{u}_k & \leq & \text{u}_\text{U}, & \forall k \in \{0,\dots, T{-}1\},
\end{array}
\end{aligned}
\end{equation}
\begin{equation}
\label{individual_ocp_common_tau_1}
\begin{aligned}
 \min_{{u}^i}&
\overbrace{\sum_{k=0}^{T-1}\Big(\| x_k^i - \tilde{\tau}^i_{k,\mathrm{ref}} \|^2_{w_x^i}+
\| u_k^i \|^2_{w_u^i}\Big)
+ \| x_T^i - \tilde{\tau}^i_{T,\mathrm{ref}} \|^2_{w_T^i}}^{J^i({u}^i; x_{\text{init}}^i, \tilde{\tau}^i_{\text{ref}}) :=} \\
\text{s.t.} &
\scalemath{0.87}{\begin{array}{rcll}
x_0^i &=& x_{\text{init}}^i, & x_{\text{init}}^i\in\mathbf{x}_\text{init} \\
h^i({x}^i_k, {u}^i_k) & = & 0, & \forall k \in \{0,\dots, T{-}1\}, \\
g_{\mathrm{h}}^{ij}(x_k^i,x_k^j,\tilde{\tau}^i_{k,\text{ref}}) &\le& 0,
& \forall k \in \{0,\dots,T{-}1\},\;
\tilde\tau^i_\text{ref} \in \tilde\tau^m_\text{top}, \\
u_{\mathrm{L}}^i \le u_k^i &\le& u_{\mathrm{U}}^i,
& \forall k \in \{0,\dots,T{-}1\},\\
\forall i,j \in {\mathcal{A}}, i \neq j
\end{array}}
\end{aligned}
\end{equation}
\revision{\noindent The reference path $\tilde{\tau}^i_{\text{ref}}$ for each agent $i \in \mathcal{A}$ is extracted from the joint candidate plan $\tilde{\tau}^m_{\text{top}}$, with the corresponding time-indexed reference denoted by $\tilde{\tau}^i_{k,\text{ref}}$ for $k = 0,\dots,T$. The vectors $\mathbf{x}$ and $\mathbf{u}$ denote the joint states and control actions, respectively, while the weights $w^i_x$, $w^i_u$, and $w^i_T$ govern state tracking, control effort, and terminal costs. The functions $h(\cdot)$ encode the joint set of equality constraints $h^i(\cdot)$ representing the system dynamics, and $g_h(\cdot)$ represents the joint set of homotopy-based collision avoidance constraints $g^{ij}_h(\cdot)$. The term $\lambda$ is a soft regularization penalizing abrupt strategy shifts. Finally, $\text{u}_\text{L}$ and $\text{u}_\text{U}$ denote the joint set of lower and upper actuator limits for each agent, $u^i_\text{L}$ and $u^i_\text{U}$.\\
$\Phi_{\text{h}}$ satisfies Definition \ref{def:wpg} by construction. Consider a unilateral deviation in control from $u^i$ to $\tilde{u}^i$ while keeping $u^{-i}$ fixed. In the difference $\Phi_{\text{h}}(u^i, u^{-i}; \mathbf{x}_{\text{init}}, \tilde{\tau}^m_{\text{top}}) - \Phi_{\text{h}}(\tilde{u}^i, u^{-i}; \mathbf{x}_{\text{init}}, \tilde{\tau}^m_{\text{top}})$, the terms $J^{-i}$ and the constant $\lambda$ cancel out, yielding:\(\Phi_{\text{h}}(u^i, u^{-i}; \cdot) - \Phi_{\text{h}}(\tilde{u}^i, u^{-i}; \cdot) = J^i(u^i; x_{\text{init}}^i, \tilde{\tau}^i_m) - J^i(\tilde{u}^i; x_{\text{init}}^i, \tilde{\tau}^i_m)\) that satisfies Definition \ref{def:wpg}}.\\
\noindent\textbf{Joint Homotopy Enforcement}
%To inform the \emph{local} game solver with the plan found by the \emph{global} homotopy planner, we formulate in \eqref{potential_nom} a joint homotopy enforcing collision avoidance constraint $g_h(\cdot)$ to simultaneously enforce the correct \emph{homotopy} class, a technique adapted from~\cite{oscar} (c.f. Fig.~\ref{fig_constraint}):$\pi_{\text{top}}^{m}$:
To inform the \emph{local} game solver with the plan from the \emph{global} homotopy planner, we formulate in \eqref{potential_nom} a joint homotopy-enforcing collision avoidance constraint $g_\mathrm{h}(\cdot)$ to enforce the correct \emph{homotopy} class, a technique adapted from~\cite{oscar} (c.f. Fig.~\ref{fig_constraint}): $\pi_{\text{top}}^{m}$:
\begin{align*}
\hat{A}^{ij}_k & := \frac{x^j_k - \tilde{\tau}^i_{k,\mathrm{ref}}}{\|x^j_k - \tilde{\tau}^i_{k,\mathrm{ref}}\|_2 + \varepsilon}, \\
\hat{b}^{ij}_k & := (\hat{A}^{ij}_k)^\top\!\left(x^j_k - \beta_o r_o \,\hat{A}^{ij}_k\right), \\
g^{ij}_{\mathrm{h}}(x^i_k,x^j_k,\tilde{\tau}^i_{k,\mathrm{ref}}) & := (\hat{A}^{ij}_k)^\top x^i_k - \hat{b}^{ij}_k, \quad \forall i,j \in \mathcal{A},~ i\neq j,\\
~k&=0,\dots,T-1.
\end{align*}
%with the compact form for every \(\tilde{\tau}^m_\text{top}\) being:
% compact form for the joint homotopy constraints
We encapsulate all $g^{ij}_{\mathrm{h}}(\cdot)$ for a given \(\tilde{\tau}^m_\text{top}\) in the compact form
\begin{equation}
g_\mathrm{h}(\mathbf{x}_k, \tilde{\tau}^m_\text{top}) \le 0, \quad  \tilde{\tau}^m_\text{top} \in \Pi_{\text{top}}.
\label{eq:compact_form}
\end{equation}
Here, $\hat{A}^{ij}_k$ is the unit normal vector (with $\varepsilon > 0$ for numerical stability) pointing from the reference path of agent $i$ toward agent $j$. The scalar $\beta_o r_o$ introduces a safety margin, ensuring that $x^i_k$ remains on the designated side of the separating hyperplane (indicated by dashed lines in Fig.~\ref{fig_constraint}) prescribed by the homotopy signature. Enforcing these constraints in \eqref{potential_nom} yields a generalized OLNE within the topologically feasible subset, ensuring local optimality while preserving the intended interaction topology.\\

\begin{figure}
\centering
\includegraphics[width=1\linewidth]{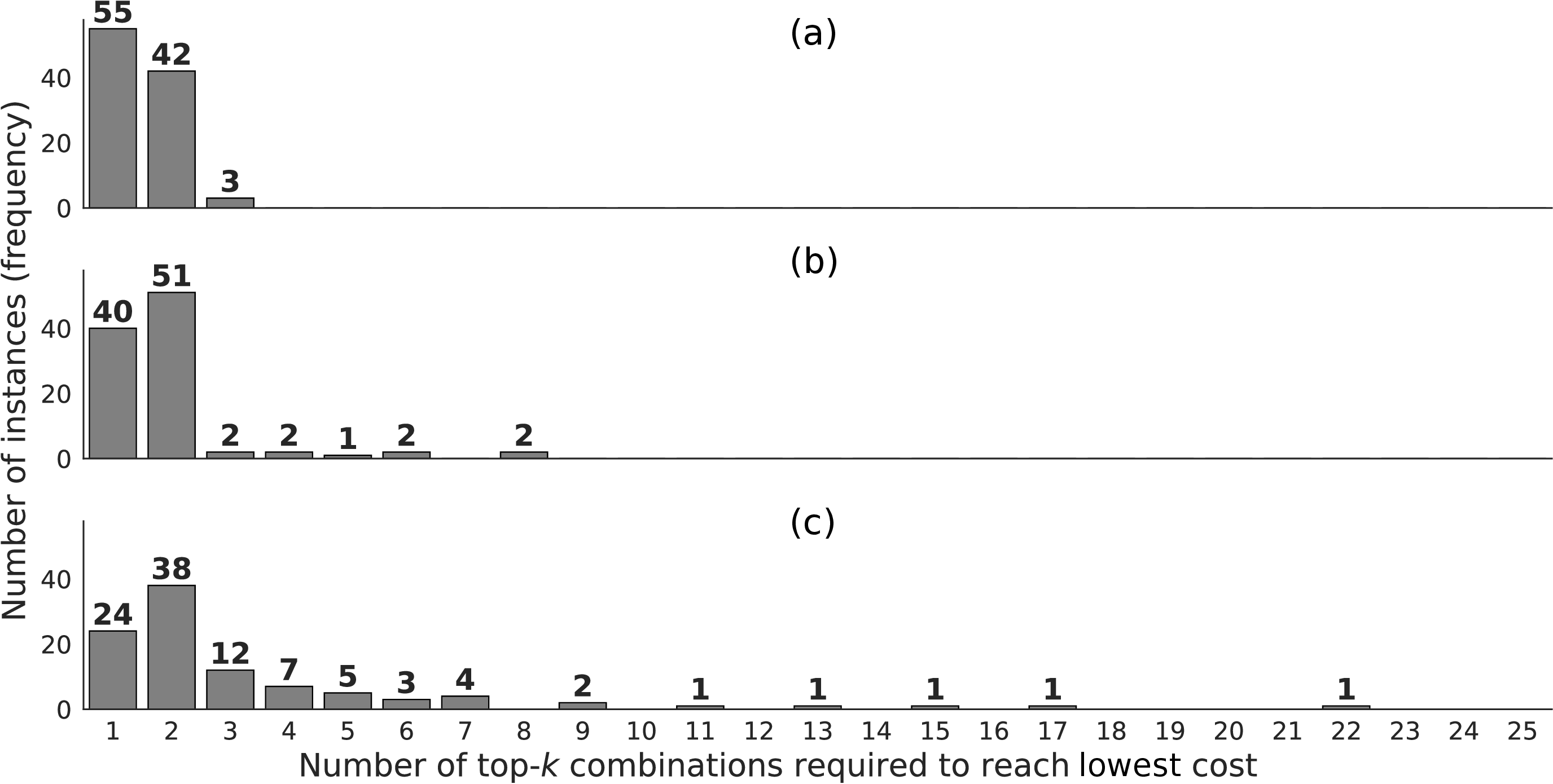}
\caption{Experiment 1: Number of top-$K$ combinations required to capture the best cost. (a) 3-player scenario, (b) 4-player scenario, (c) 5-player scenario.}
\label{comp_plot_decay}
\end{figure}

\revision{\noindent\textbf{Experiment 1: Heuristic Prefilter Performance Evaluation} 
To manage combinatorial complexity, the heuristic prefilter selects $K$ promising candidates $\Pi_\text{top} \subseteq \Pi_\text{valid}$.
However, the candidate strategy ranked highest in $\Pi_\text{top}$ according to our heuristic may not yield the lowest \emph{true} cost of~\eqref{potential_nom}, necessitating the evaluation of $K > 1$ candidates to find the best strategy.
Experiment~1 evaluates the trade-off between computational efficiency and solution accuracy, motivating the choice of $K$. To verify that $\Pi_\text{top}$ reliably contains the lowest game cost combination, 100 simulations were run for 3-, 4-, and 5-player scenarios. Start and goal positions were selected to generate a rich set of valid homotopy combinations $\Pi_\text{valid}$. The heuristic filter was disabled, and \eqref{potential_nom} was solved in open loop for 4~s (40 steps) over all homotopy combinations. For each trial, the minimum-cost combination was identified, and its rank within the prefilter's list was recorded. Figure~\ref{comp_plot_decay} shows histograms of these ranks. The maximum rank with non-zero frequency gives a lower bound on $K$ needed to reliably capture the lowest-cost candidate combination. As \(N\) increases, $K$ grows, empirically within the 3rd, 8th, and 22nd ranked combinations for 3-, 4-, and 5-player cases. These results give an estimate for $K$ and show that the lowest cost can be achieved empirically with a substantially reduced search space.}

\noindent\textbf{Homotopy Consistency in Receding-Horizon Invocations.}  
\revision{
%\linelabel{sec_3_nov}
To avoid abrupt strategy shifts in receding-horizon implementations, we add a soft penalty $\lambda$ in~\eqref{potential_nom}, defined as
\begin{equation}
\label{lambda_penalty}
\lambda = \sum_{i=1}^{N} 
\mathbb{1}_{\left\{ \mathcal{H}_\text{p}^i \in \Pi_{\text{top}} \;\land\; \mathcal{H}_\text{c}^i \neq \mathcal{H}_\text{p}^i \right\}} \cdot \lambda_{\text{pen}},
\end{equation}
where $\lambda_\text{pen}$ is a scalar constant. This term preserves the symmetry of the underlying potential game while favoring the previously selected homotopy class $\mathcal{H}_\text{p} = \{ \mathcal{H}_\text{p}^i (\hat{\tau}^i) \}_{i=1}^N$, with $\hat{\tau}^i$ being the reference trajectory of agent $i$ extracted from the best top-ranked plan at the previous instant. The preference is maintained unless a newly available homotopy configuration $\mathcal{H}_\text{c} = \{ \mathcal{H}_\text{c}^i (\hat{\tau}^i) \}_{i=1}^N$ yields a lower overall cost after including the penalty, or if the previously selected homotopy is no longer feasible.}
%\linelabel{sec_3_nov_end}
\revision{
%\linelabel{lamb_ex}
This allows the solution to adapt, switching from initial paths that may become inefficient to better homotopy combinations while maintaining stability through the regularization of $\lambda$.
%\linelabel{lamb_ex_end}
%\linelabel{lamb_0}
Smaller \(\lambda\) values enable easier switching, leading to more frequent cumulative homotopy changes, while larger \(\lambda\) values favor persistence of the chosen combination
%\linelabel{lamb_1}
} \revision{\footnote{Even though our approach explores solutions in the global space, due to the non-convex nature of the formulation, global optimality cannot be stated.}}.
\section{Simulation Results}
\label{sec:sim_results}
\begin{figure}[t]
\centering
\includegraphics[width=1\linewidth]{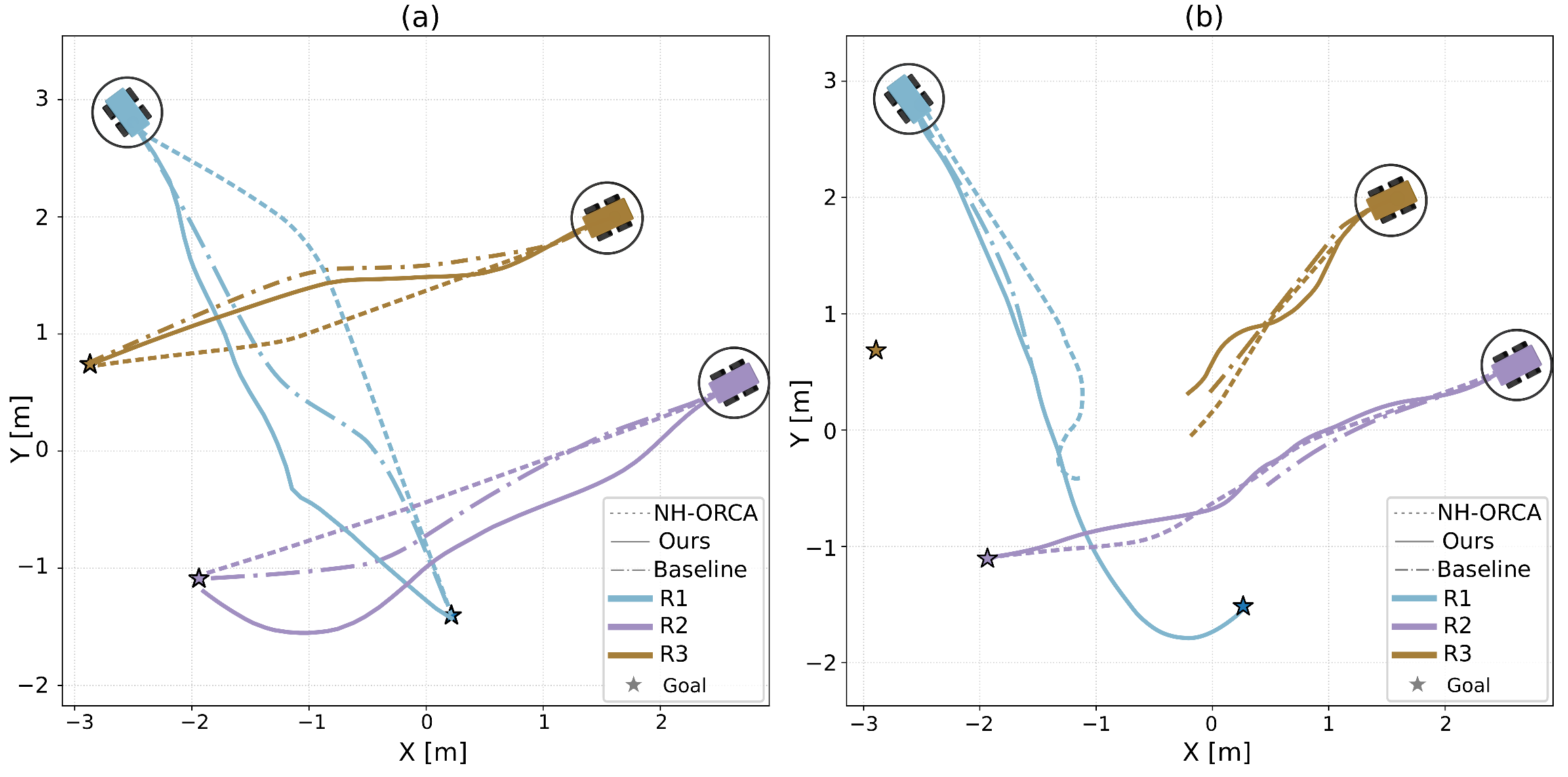}
\caption{Comparison between homotopy-assisted, baseline potential game formulations, and NH-ORCA over 100 evaluated scenarios with agents R1, R2, and R3. (a) Robot trajectories for the representative sample scenario under rational agent setting; (b) Robot trajectories for the representative sample scenario under irrational agent }
\label{comp_plot_bin}
\end{figure}

We evaluate our approach in simulation using a differential-drive robot in a three-player game. The robot follows standard unicycle kinematics: 
\(\dot{x} = v \cos\theta, \; \dot{y} = v \sin\theta, \; \dot{\theta} = \omega\), 
with control bounds \(v \in [\SI{0}{\meter\per\second}, \SI{0.55}{\meter\per\second}]\) and \(\omega \in [\SI{-0.5}{\radian\per\second}, \SI{0.5}{\radian\per\second}]\). Cost weights are \(w_x = w_y = 100\), \(w_v = 1\), \(w_\omega = 30\), and for homotopy-based reasoning we set \(\beta_o = 1\), \(r_o = \SI{1}{\meter}\), and \(\lambda_\text{pen} = 100\). As shown in Fig.~\ref{comp_plot_decay}, the top three combinations from prefilter consistently capture the lowest cost of \eqref{potential_nom} in the three-player case. To improve robustness, we use \(K=4\) for top-\(K\) selection. The receding-horizon parameters are \(T=40\) and \(\Delta_\text{step} = \SI{0.1}{\second}\). \revision{We benchmark our approach against the following frameworks, which does not reason about homotopies:

\noindent\textbf{NH-ORCA:} The velocity-space NH-ORCA model~\cite{AlonsoMora2010OptimalRC} uses a robot radius
\(r_{i,\mathrm{ORCA}} = \SI{0.45}{\meter}\) with a tracking error bound
\(\mathcal{E} = \SI{0.05}{\meter}\), giving an effective radius
\(r_{\mathrm{eff}} = r_{i,\mathrm{ORCA}} + \mathcal{E} = \SI{0.5}{\meter}\).  
Reference velocities \((\dot{x}^i_\mathrm{ref}, \dot{y}^i_\mathrm{ref})\) point from each agent’s position toward its goal.  
%Non-holonomic velocities are mapped to admissible holonomic velocities using an offline polytope computed from \(\mathcal{E}\) and \(T_{t1} = \SI{0.5}{\second} \) in the formulation.
Non-holonomic velocities are mapped to admissible holonomic velocities using an offline polytope in the setup to ensure that the resulting velocities remain feasible.\\
\noindent\textbf{Baseline Potential Game:}
The baseline potential game model uses shortest straight-line reference paths $\tilde{\tau}^i_{\mathrm{b}}$ between each agent’s start and goal positions and enforces a minimum inter-agent separation of $\SI{1}{\meter}$ via $\tfrac{1}{2}N(N-1)$ pairwise Euclidean distance constraints\(\| x_k^i - x_k^j \|_2^2 \ge 1^2,\; \forall i,j \in \mathcal{A},\; i<j,\; k=0,\dots,T{-}1.\) These constraints are encapsulated in $g_{\mathrm{b}}(\cdot)$ in~\eqref{potential_BASE}. All simulations are implemented in \texttt{acados} using an SQP solver~\cite{Verschueren2021}. Receding horizon parameters remain the same for our approach and baseline game.
\begin{equation}
\label{potential_BASE}
\begin{aligned}
\min_{\mathbf{u}}~ & 
\overbrace{
\scalemath{0.87}{
\sum_{k=0}^{T-1} \sum_{i=1}^{N} 
\left( \| x^i_k - \tilde{\tau}^i_{k,\mathrm{b}} \|^2_{w^i_x} + \| u^i_k \|^2_{w^i_u} \right)
+ \sum_{i=1}^{N} \| x^i_T - \tilde{\tau}^i_{T,\mathrm{b}} \|^2_{w^i_T}
}
}^{\Phi_\mathrm{b}(\mathbf{u}; \mathbf{x}_\mathrm{init}) :=}
\\
\text{s.t.}\quad & 
\begin{array}{rcll}
\mathbf{x}_0 & = & \mathbf{x}_\mathrm{init}, & \forall i \in \mathcal{A}, \\
h(\mathbf{x}_k, \mathbf{u}_k) & = & 0, & \forall k \in \{0,\dots, T{-}1\}, \\
g_\mathrm{b}(\mathbf{x}_k) & \le & 0, & \forall k \in \{0,\dots, T{-}1\}, \\
\mathbf{u}_\mathrm{L} \le \mathbf{u}_k & \le & \mathbf{u}_\mathrm{U}, & \forall k \in \{0,\dots, T{-}1\}.
\end{array}
\end{aligned}
\end{equation}
\noindent\textbf{Results and Discussion.}  
We simulate 100 randomized scenarios under a full-information setup for all three frameworks. 
Results are listed in Table~\ref{tab:combined_results_manu} using performance and comparative metrics, with representative trajectories shown in Fig.~\ref{comp_plot_bin}(a).\\% 
\textbf{Performance Metrics:} From the performance metrics, our approach achieves faster mean completion time (\(+9.41\%\) vs. Baseline, \(+9.86\%\) vs. NH-ORCA) at the cost of slightly longer paths (\(+2.83\%\) vs. Baseline, \(+1.62\%\) vs. NH-ORCA), while yielding larger inter-agent distances (\(+22.8\%\) vs. Baseline, \(+28.1\%\) vs. NH-ORCA) by reasoning over multiple homotopies and applying the heuristic prefilter.\\
\textbf{Comparative Metrics:} Comparative Outcomes quantify per-instance wins for shortest path and fastest completion time, which are key objectives in navigation tasks, across the frameworks. A "win" identifies the best-performing framework for that specific objective in each task instance, reflecting interaction efficiency. For a 3-agent game over 100 scenarios, there are \(3 \times 100 = 300\) agent instances. Each agent is evaluated under the three frameworks, and each case receives 1 point if it achieves a shorter path or faster completion than the other cases for that agent instance. Under our approach, 171 agent instances complete faster (\(+470\%\) relative to Baseline), while 62 achieve the shortest path (\(+8.77\%\)). NH-ORCA achieves 181 shortest-path wins (\(+217.5\%\)) and 99 fastest-time wins (\(+230\%\)), whereas the Baseline achieves 57 shortest-path wins and 30 fastest-time wins. Overall, our method prioritizes faster completion while maintaining enhanced safety, as reflected by larger inter-agent distances, validating our first two claims.\\
\noindent\textbf{Irrational Setting:} In the scenario of Fig.~\ref{comp_plot_bin}(a), Agent R3 is made irrational to evaluate robustness. As shown in Fig.~\ref{comp_plot_bin}(b), both the Baseline and NH-ORCA fail: the Baseline cannot generate alternative local solutions under irrational behavior, while NH-ORCA assumes shared collision-avoidance responsibility from R3, which does not occur, causing agent R1 to become trapped between R2 and R3. In contrast, our approach safely navigates around R3 by generating alternative homotopy candidates, effectively synthesizing feasible local solutions.}
\begin{table}
\centering
%\caption{Quantitative evaluation of navigation performance and competitive outcomes across 100 trials.}
\caption{Quantitative evaluation of navigation performance and comparative outcomes across 100 trials.}
\label{tab:combined_results_manu}
\footnotesize
\setlength{\tabcolsep}{3pt}
\renewcommand{\arraystretch}{1}
\begin{tabular}{p{0.16\columnwidth} p{0.22\columnwidth} p{0.26\columnwidth} p{0.24\columnwidth}}
\hline
\multicolumn{4}{c}{\textbf{Performance Metrics (Mean Values)}} \\
\hline
\textbf{Case} 
& \parbox[c]{0.24\columnwidth}{\centering\textbf{Path Length}\\\textbf{[m]}} 
& \parbox[c]{0.24\columnwidth}{\centering\textbf{Min}\textbf{Inter-Agent Dist [m]}} 
& \parbox[c]{0.24\columnwidth}{\centering\textbf{Completion}\\\textbf{Time [s]}} \\
\hline
Ours     
& $5.81 \pm 0.94$ ($+2.83\%$) 
& $2.53$ ($+22.8\%$) 
& $16.06 \pm 1.89$ ($-9.41\%$) \\
Baseline 
& $5.65 \pm 0.82$ 
& $2.06$ 
& $17.73 \pm 1.27$ \\
NH-ORCA  
& $5.73 \pm 0.99$ ($+1.42\%$) 
& $1.95$ ($-5.34\%$) 
& $17.81 \pm 2.82$ ($+0.45\%$) \\
\hline
\multicolumn{4}{c}{\textbf{Comparative Outcomes (Wins) }} \\
\hline
\textbf{Case} 
& \parbox[c]{0.22\columnwidth}{\centering\textbf{Shortest Path}} 
& \parbox[c]{0.26\columnwidth}{\centering\textbf{Fastest Time}} 
&  \\
\hline
Ours     
& 62 ($+8.77\%$) 
& 171 ($+470\%$) 
&  \\
Baseline 
& 57 
& 30 
&  \\
NH-ORCA  
& 181 ($+217.5\%$) 
& 99 ($+230\%$) 
&  \\
\hline
\multicolumn{4}{l}{\footnotesize Percentage differences are reported relative to the baseline method.} \\
\end{tabular}
\end{table}

\begin{table}[ht]
\centering
\caption{Computation time breakdown for $N=3$--$6$ agents. Times are reported as mean $\pm$ standard deviation [ms].}
\label{computation_time_N3_N6}
\footnotesize
\setlength{\tabcolsep}{2pt}
\renewcommand{\arraystretch}{1}
\begin{tabular}{lcccc}
\hline
\textbf{Stage} & \textbf{$N=3$} & \textbf{$N=4$} & \textbf{$N=5$} & \textbf{$N=6$} \\
\hline
Homotopy Planner      & $16 \pm 2$      & $20 \pm 3$      & $30 \pm 4$      & $337 \pm 172$ \\
Heuristic Prefilter   & $0.35 \pm 0.20$ & $1.5 \pm 1.8$   & $21 \pm 31$     & $624 \pm 1859$ \\
\hline
GT-Solver SQP-RTI     & $6 \pm 3$       & $22.7 \pm 10.5$ & $117 \pm 44$    & $183 \pm 16$ \\
Total SQP-RTI         & $22.35 \pm 3.9$ & $44.2 \pm 11.6$ & $168 \pm 55.4$  & $1144 \pm 2046$ \\
\hline
GT-Solver SQP         & $28 \pm 17$     & $52 \pm 31$     & $181 \pm 50$    & $308 \pm 62$ \\
Total SQP             & $44.35 \pm 17.2$& $73.5 \pm 31.4$ & $232 \pm 57$    & $1269 \pm 2047$ \\
\hline
\end{tabular}
\end{table}
\begin{figure*}
   \centering
 \subfloat{%
      \includegraphics[width=0.5\linewidth]{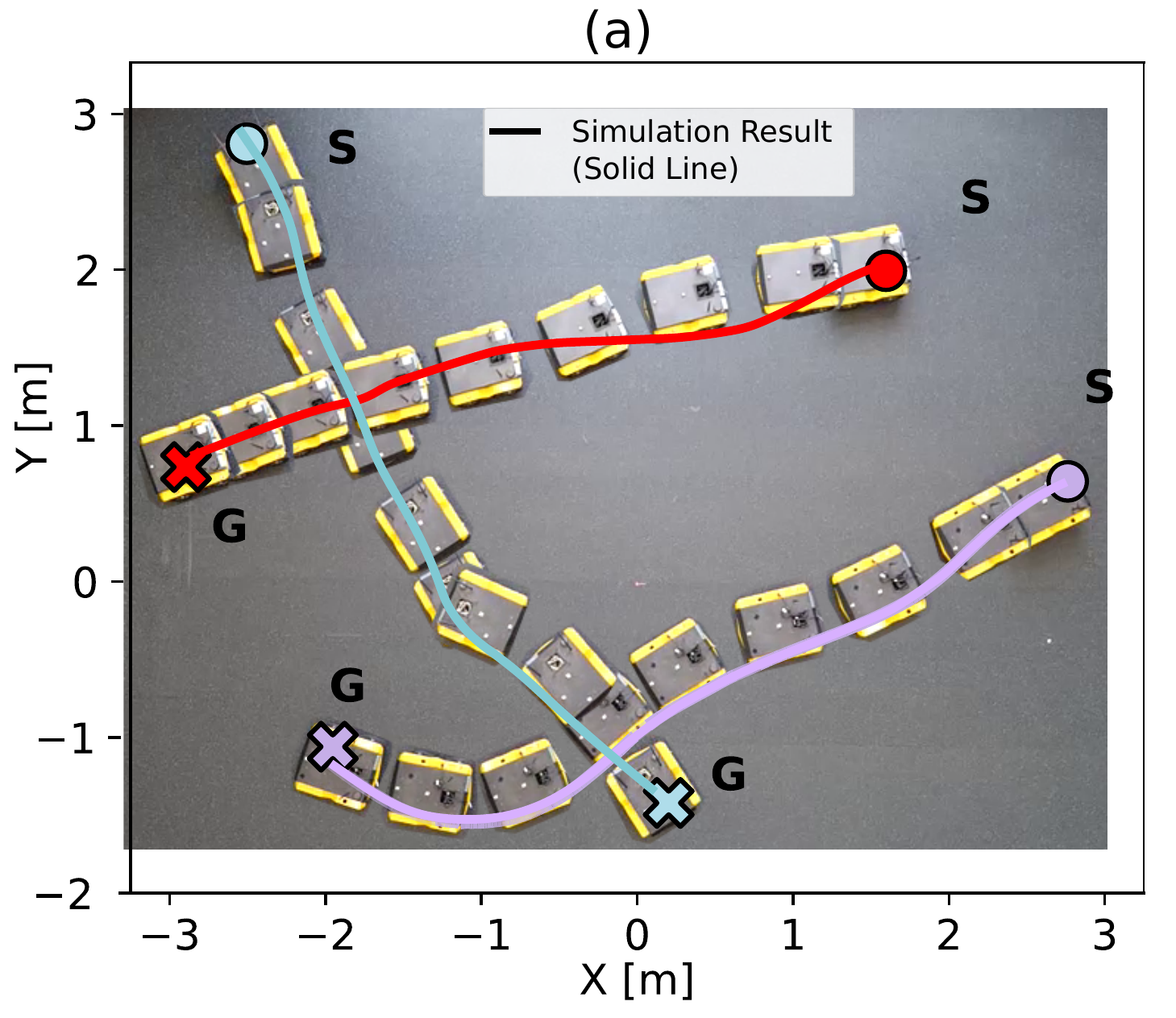}}
      \hfill
\subfloat{%
       \includegraphics[width=0.5\linewidth]{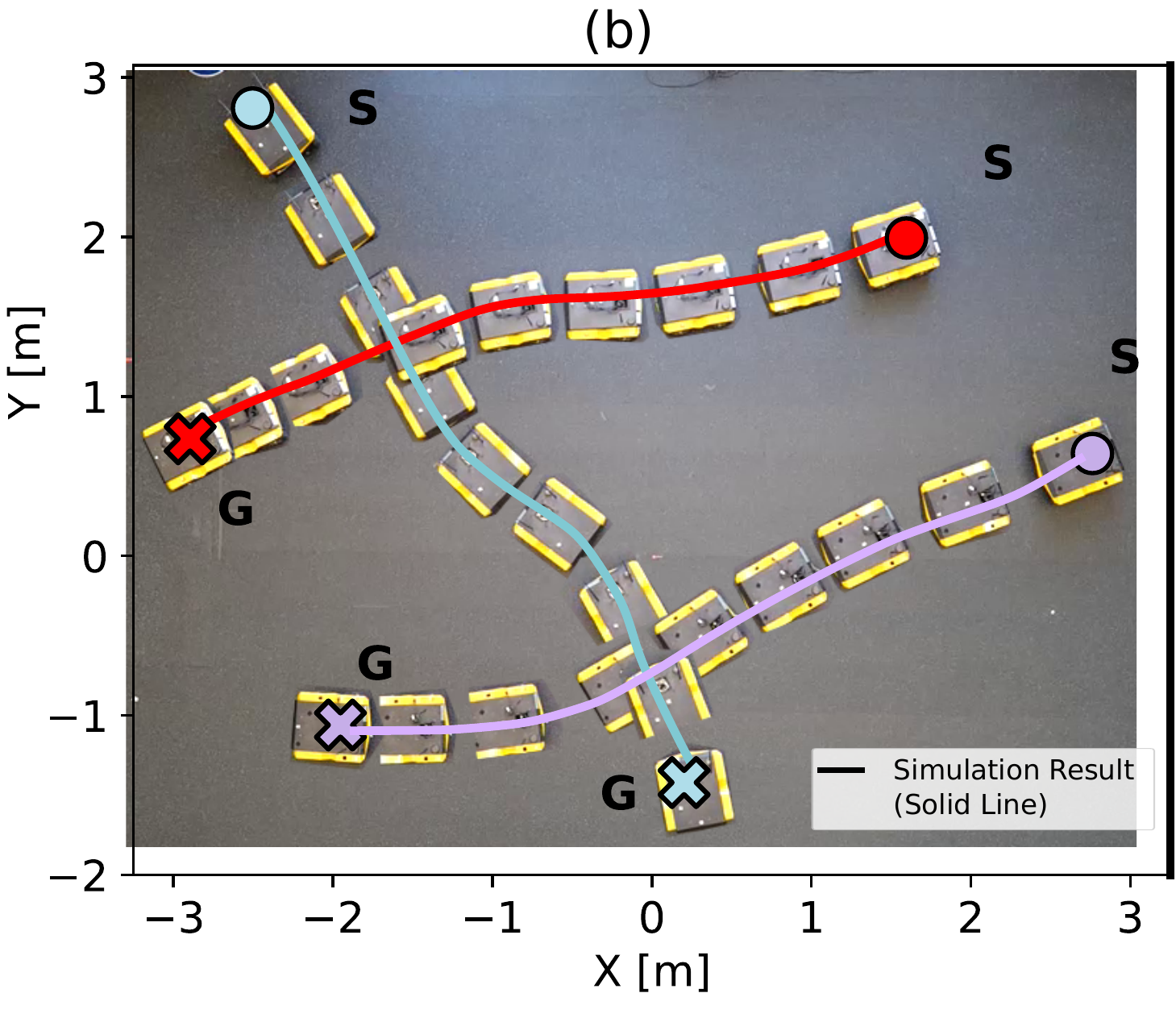}}
             \hfill
 \subfloat{%
       \includegraphics[width=0.5\linewidth]{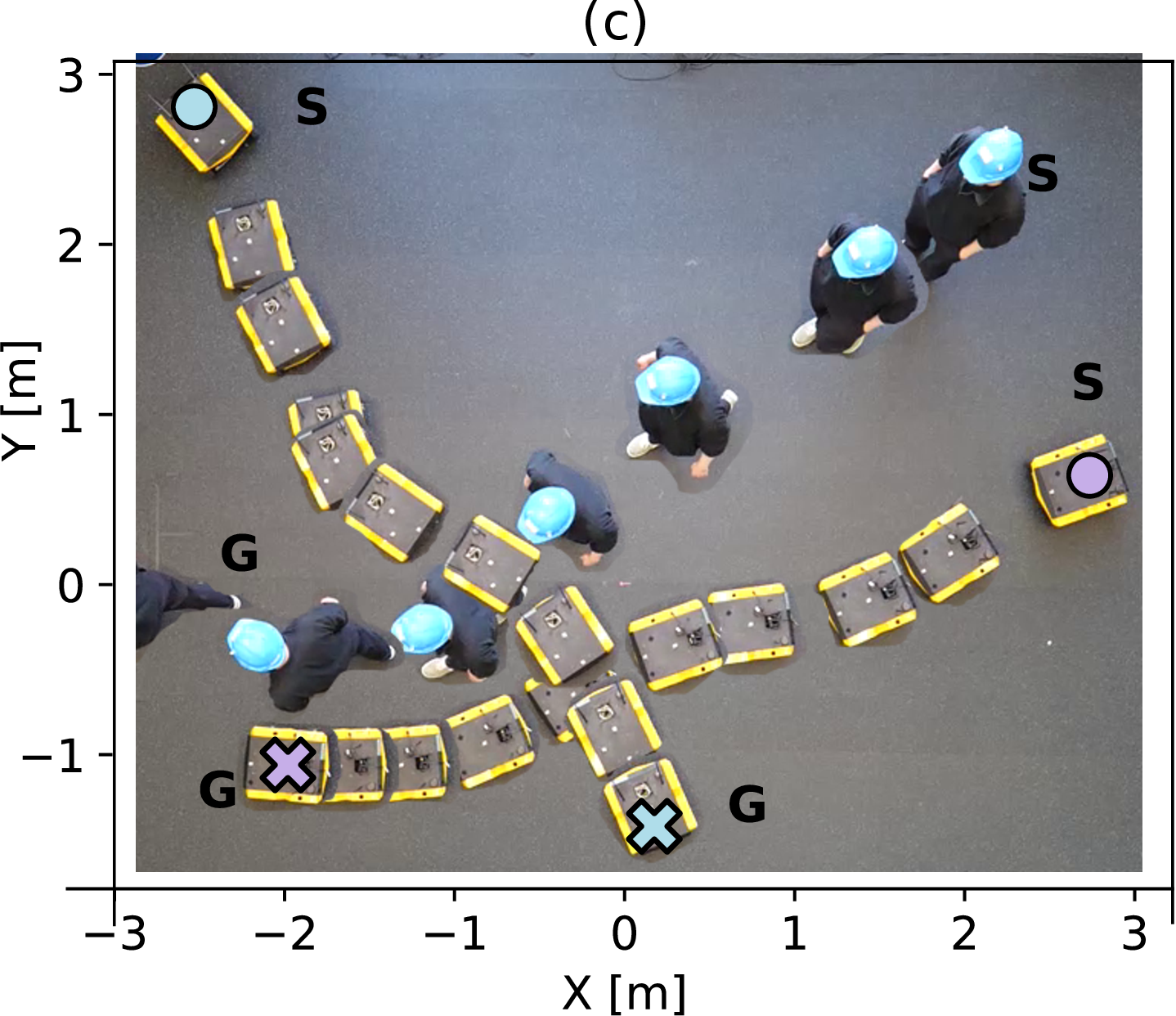}}
       \hfill
 \subfloat{%
       \includegraphics[width=0.5\linewidth]{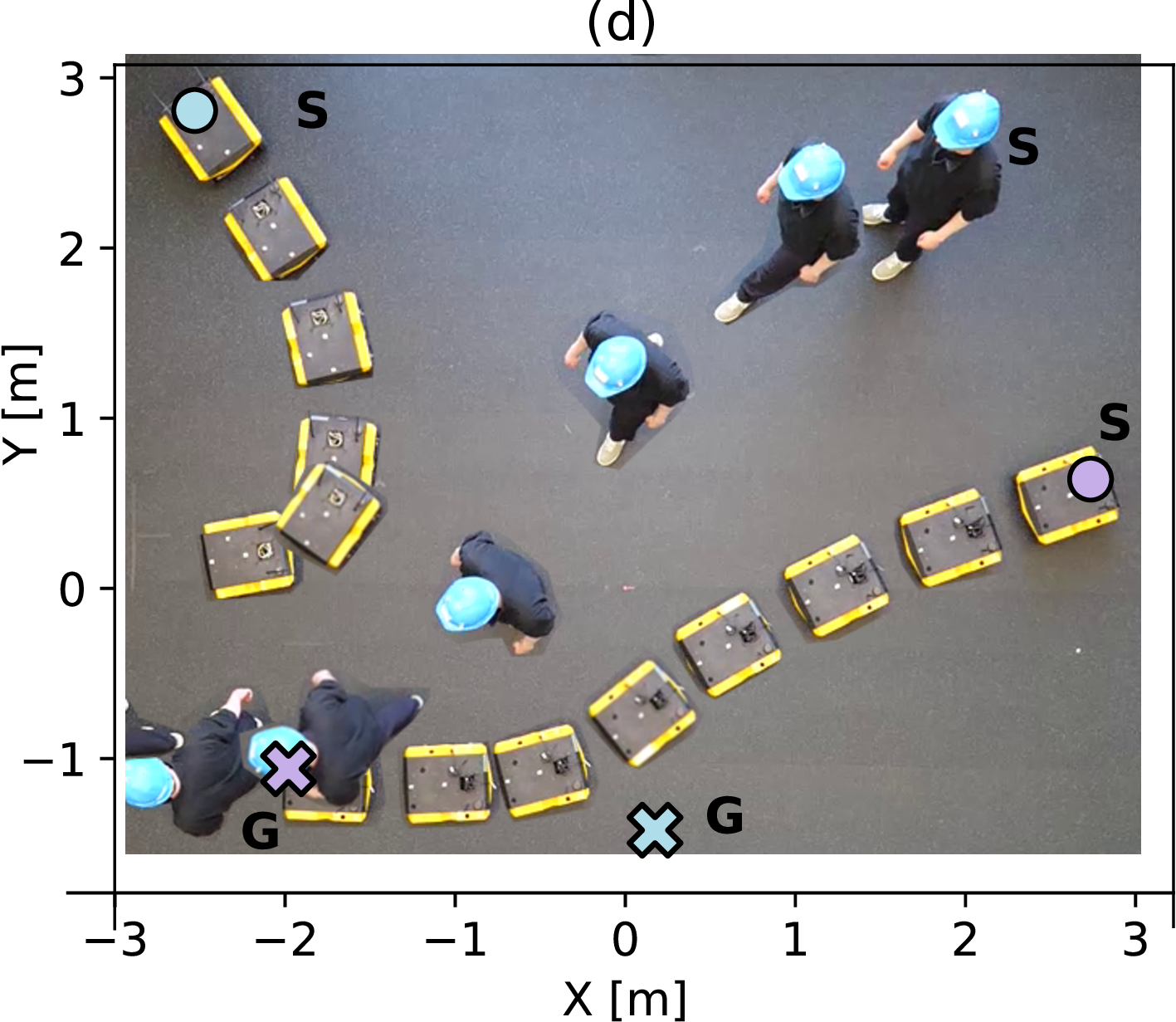}}
        
    \caption{Hardware validation results for our approach and the baseline game under rational and irrational settings for a three-player game (Agents R1, R2, R3), shown in light blue, light pink, and red. \(\textbf{S}\) and \(\textbf{G}\) represent Start and Goal respectively.
   (a) Rational scenario under the proposed approach, with the corresponding simulation output superimposed for comparative analysis(\href{https://youtu.be/A1dnBF\_SUxM}{see video (V1)}). 
    (b) Rational scenario under the baseline approach with no homotopy reasoning, with the corresponding simulation output superimposed for comparative analysis(\href{https://youtu.be/OUM2H7aAL7E}{see video (V2)}). 
    (c) Irrational scenario under the proposed approach (\href{https://youtu.be/2u1zIdy9ZTg}{see video (V3)}). 
    (d) Irrational scenario under the baseline approach with no homotopy reasoning (\href{https://youtu.be/U-8_nv9J3Zk}{see video (V4)}).}
    \label{fig:comp_plot_2x2}
\end{figure*}
\section{Hardware Validation}
\label{sec:hardware_validation}
We validated the proposed framework against the baseline game on physical Clearpath Jackal robots in two scenarios. % ADD A NOTE ON ACADOS%
In both cases, initial and goal positions matched those used in the representative simulation to ensure consistency. The frameworks were implemented online on a central computer, with control inputs transmitted to the robots via WebSockets and perception provided by a VICON motion capture system. \\
\textbf{Scenario 1: Rational Agents.} In this scenario, all agents followed their objectives rationally. Our homotopy-guided framework (Fig.~\ref{fig:comp_plot_2x2}(a)) and the baseline (Fig.~\ref{fig:comp_plot_2x2}(b)) produced trajectories consistent with the simulation results in a topological sense, confirming reproducibility on physical hardware.\\
\textbf{Scenario 2: Irrational Agent.} In this scenario, one Jackal robot is replaced by a human pedestrian who behaves unpredictably, not following equilibrium-consistent behavior. Our framework enables agents find an alternative local NE, by bypassing the pedestrian while maintaining feasibility (Fig.~\ref{fig:comp_plot_2x2}(c)). In contrast, the baseline method (Fig.~\ref{fig:comp_plot_2x2}(d)) fails to adapt, causing a solver crash. These results confirm our third claim: our framework remains robust to irrational agent behavior.
\section{Scalability Analysis}
\label{sec:scalability}
\revision{
To evaluate scalability, we conducted 100 randomized simulations for 3-, 4-, 5-, and 6-player games, recording the total per-step solution time (sum of homotopy planning, heuristic prefiltering, and SQP/SQP-RTI solver setups) as reported in Table~\ref{computation_time_N3_N6}. The results indicate that directly scaling the approach beyond $N \gg 5$ becomes intractable, motivating the use of techniques such as~\cite{10341677}. Reducing the computational burden at the homotopy planning stage remains an ongoing direction for dense-agent settings, for instance, by leveraging hyperplane-based space subdivision to focus homotopy construction on the most promising regions.
}

\section{Conclusion}
In this paper, we explored the complementary benefits of combining homotopy planning with game-theoretic modeling for multi-agent motion planning under congestion. Simulation results showed that our framework outperformed both the baseline game and NH-ORCA, which do not reason about homotopies, in scenario completion time, inter-agent distance (indicating reduced congestion), with only a marginal increase in path length. Hardware validation further demonstrated robustness to irrational agent behavior in a 3-player game. \\
%\linelabel{limi_start}
However, the framework relies on simplifying assumptions, including full state observability, fixed planning horizons, and static agents during homotopy planning. These assumptions may constrain performance in more complex and densely populated environments. %Future work will focus on extending the approach to three-dimensional strategy spaces and addressing asymmetric agent behavior beyond potential game formulations. 
\revision{Future work will focus on relaxing many of these assumption and addressing asymmetric agent behavior beyond potential game formulations.}
%\linelabel{limi_start_end}

\bibliographystyle{IEEEtran}
\bibliography{IEEEabrv, references}
\end{document}